# Electron-electron attraction via Coulomb correlations and possible superconductivity in a 1D electron liquid on a rigid neutralizing background

Yu. P. Monarkha

*B. Verkin Institute for Low Temperature Physics and Engineering of the National Academy of Sciences of Ukraine Kharkiv 61103, Ukraine*

E-mail: monarkha@ilt.kharkov.ua

Conditions at which a quasi-one-dimensional (1D) electron system can be considered as a quantum liquid of impenetrable charged particles are theoretically analyzed. In the presence of an inert, neutralizing background, a motion of impenetrable electrons is shown to expose a positive charge, resulting in an effective mutual attraction of infinite range. As a result, all electrons are involved in the long-range pairing. A model of spinless fermions with infinitesimal attraction of infinite range is proposed to describe the excitation spectrum and the superconducting gap in low-density 1D electron channels. In contrast with the conventional theory, the energy gap does not contain exponentially small factors. It depends mostly on the Coulombic parameters of the system, which guides practical aspects of high-temperature superconductivity.

## 1. Introduction

The discovery (Onnes, 1911) and understanding [Bardeen–Cooper–Schrieffer (BCS) theory, 1957] of the superconductivity phenomenon represent the most outstanding achievements in quantum solid-state physics. Despite its universality proved for many materials, superconductivity remains a low-temperature phenomenon. Nevertheless, there are suppositions that, under suitable conditions, superconductivity can occur above room temperature as well (for a review, see Ref. [1]). The basic ideas towards a substantial increase of a transition temperature $T_c$ discussed in the literature relate mostly to a modification of Cooper pairing and low-dimension physics [1, 2].

The attractive force responsible for Cooper pairing is caused by a disturbance of the ion lattice produced by a moving electron. The area of higher positive charge density attracts another electron, which can be described as a weak electron-electron interaction. For conventional Cooper pairs, the disturbance of ion lattice is very small because of the significant difference in mass between electrons and ions. Additionally, other electrons screen fast a charge imbalance. Remarkably, even Coulomb repulsion between electrons in usual Fermi-liquid can produce a weak attraction and superconductivity [3]. The physics of this effect is similar to the well-known Friedel oscillations of electron density accompanying the external charge screening. The effective interaction between the fermions themselves has a long-range oscillatory part, and Cooper pairs are formed due to the attractive regions [3]. Still, the critical temperature for this mechanism is estimated to be exponentially small.

For the Friedel oscillations, the attractive regions are very small, because they are defined by the Fermi momentum $k_F$: $\delta x \sim 1/k_F$. Amazingly, in a quasi-one-dimensional (1D) electron system confined to a narrow channel with a uniform (along the channel), inert, neutralizing background, under certain conditions, the positive charge can be strongly exposed by simple electron motion, forming a wide attractive region [4]. In this case, an attractive potential affects neighboring electrons within a macroscopically long range. Such conditions appear in a low density limit, when the electron Fermi energy $\mathcal{E}_F$ is substantially smaller than the Coulomb potential at small distances $V_d = e^2/d$ (here $d$ is the channel width, assumed to be about the atomic size and much smaller than the typical electron spacing $a = 1/n_e$). Therefore, a moving electron cannot pass by a neighboring electron (penetrate the other side), and involves it in the motion. Under these conditions, the electrons can be considered as impenetrable particles. Nevertheless, if electron density $n_e$ is low ($a \gg d$), but $\mathcal{E}_F$ is still much larger $e^2/\varepsilon a$ (here $\varepsilon$ is the static dielectric constant of the medium) electrons cannot be localized within the spacing $a$ because of quantum effects: otherwise each localized electron would have the kinetic energy $\simeq \hbar^2/2m^*a^2 \approx \mathcal{E}_F$ (here $m^*$ is the effective mass of electrons).

That is, the Wigner solid or a classical liquid cannot be formed in the channel under the above-noted conditions because the amplitude of zero-point vibrations is much larger than $a$. It is well-known that even bosons having impenetrable core can be described as spinless fermions [5], i.e., free particles. Moreover, the system represents a rather intriguing 1D quantum liquid, where electron motion exposes the positive background to a large extent without screening, which is the reason for a long-range attraction between electrons. Thus, in such a liquid, it is possible to avoid the two important negative factors reducing $T_c$ in conventional superconductors: strong screening of attractive regions and the heavy restriction on the number of electrons involved in pairing (only about $10^{-4}$ fraction). Regarding the relationship between the present study and the well-known results obtained for 1D systems [6] using the perturbation theory, it is important to note that here we intend to describe a system with extremely strong interaction which imposes certain symmetry on the many-electron wavefunction and leads to attractive potentials of infinite range.

In this work, we develop the theoretical description of the quantum quasi-1D liquid of impenetrable electrons formed on a rigid, uniform, positive background. Using a mean field approximation, a complex many-electron effect produced by the background exposed is reduced to a two-particle attraction acting on the nearest neighbors. The long-range attractive potential is responsible for phase transitions in such a 1D system, which is clearly shown here for classical and quantum limits. To investigate the quantum properties of low-density 1D liquid, we propose a model of spinless fermions with infinitesimal attractive force of infinite range. Its advantage is the possibility of using theoretical methods that proved to be effective for the description of conventional superconductivity. It is important that the quasiparticle energy gap found here has no exponentially small factors and gives optimistic estimations for the critical temperature.

## 2. Effective interaction potential

Consider a model quasi-1D system illustrated schematically in Fig. 1. Electrons are confined within a narrow positively charged channel of a microscopic diameter $d$. For convenience, the figure has different scales along the channel and in the transverse direction. The background charge of density $n_p$ is assumed to be rigid, uniform (along the channel), and inert. In most cases, we shall consider that electron density $n_e = n_p$, and the electron spacing $a \equiv 1/n_e$ is much larger than the channel width $d$. At small distances between two electrons $x \to d$, the Coulomb interaction energy $V_d \simeq e^2/d$ is not affected by any macro-scopic screening and, therefore, is high. We assume $V_d$ is the highest energy parameter in the system. At larger distances $x \gtrsim a$, the media screening is taken into account by employing a static dielectric constant $\varepsilon$ which could be rather high ($V = e^2/\varepsilon x$). Thus, at average distances $x \sim a$, the direct Coulomb interaction energy of a pair of electrons is rather weak, but it becomes very strong when $x$ approaches $d$.

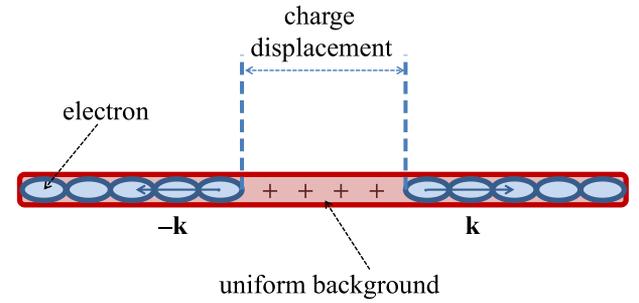

*Fig. 1.* (Color online) Schematic illustration of a quasi-1D electron liquid formed on a rigid, uniform, positive background. Electrons are impenetrable because of strong Coulomb interaction at small distances. Therefore, the opposite motion of the two nearest electrons exposes the background charge, forming an attractive region.

The Fermi energy $\mathcal{E}_F$ is another important parameter of the model proposed here. In the following, we shall consider electrons with the spin number $s = 1/2$ and a model of spinless fermions ($s = 0$). Therefore, for a 1D system, the Fermi energy and momentum are defined as

$$\mathcal{E}_F = \frac{1}{2_s^2}\frac{\hbar^2\pi^2 n_e^2}{2m^*}, \quad k_F = \pi n_e/2_s, \qquad (1)$$

where $2_s$ equals 2 if $s = 1/2$ and 1 if $s = 0$. Thus, the conditions necessary to realize the quasi-1D quantum liquid of impenetrable electrons can be formulated as

$$\frac{e^2}{\varepsilon a} \ll \mathcal{E}_F \ll \frac{e^2}{d}. \qquad (2)$$

This is why we assume a low electron density ($a \gg d$) and substantial static screening. The effective electron mass $m^*$ can also help with tuning the system to the conditions of Eq. (2).

In the Introduction, we already emphasized that at $\mathcal{E}_F \gg e^2/\varepsilon a$ the amplitude of zero-point vibrations of electrons is much larger than their average spacing $a$. Therefore, individual shapes of electron clouds shown in Fig. 1 should be considered only as illustration of the average electron spacing. The quantum electron liquid is smooth. A charge displacement produced by electron motion is shown in this figure by horizontal arrows. In the system of impenetrable electrons, the electron displacement exposes the positive background which attracts back not only the nearest electrons but also the all other electrons from both sides of the channel. The attractive force acting on other electrons of the electron line (left or right) eventually is applied to the edge electron.

For a given displacement amplitude $x$, the attractive force acting on electrons of the right line can be represented as

$$F_R(x) = \frac{n_e n_p e^2}{\varepsilon} \int_x^\infty \int_{a/2}^{x-a/2} \frac{1}{(x_e - x_p)^2} dx_p dx_e$$

$$= \frac{n_e n_p e^2}{\varepsilon} \ln\left(2\frac{x}{a} - 1\right). \quad (3)$$

Here, we disregarded the finite thickness of the channel as compared with $a$ and assumed that electron density is approximately constant along the shifted lines (increasing density would give stronger force). It should be noted that $F_R(x) \to 0$ at equilibrium ($x = a$), and it increases slowly (logarithmically) with $x$. At this point, it should be noted that there is a similar attractive force acting between other electrons (not only between nearest neighbors) with a reduced background charge exposed, but it will be disregarded in the following treatment.

Employing Eq. (3), the potential energy obtained by separation ($x > a$) of two nearest neighbors can be found as

$$V(x) = V_B \varphi(x/a) \theta(x/a - 1), \quad (4)$$

where $V_B = e^2 n_p / \varepsilon$ is the potential parameter due to the background, $\theta(x)$ is the Heaviside step function, and

$$\varphi(\rho) = \left(\rho - \frac{1}{2}\right) \ln(2\rho - 1) - (\rho - 1). \quad (5)$$

From Eqs. (4) and (5), it follows that the potential energy of the two nearest electrons increases with $x$ even stronger than a linear function [in our short note [4], we used only the approximation $V(x) \propto x$]. For smaller distances $a > x > d$, the dependence $V(x)$ can be approximated by $e^2/\varepsilon x$; in the following, we shall see that this range is less important for our goals. In this work, we consider only the case of zero pressure $P = 0$. Generally, a finite $P$ would lead to an additional term $P \cdot (x - a)$ in Eq. (4).

The potential energy of a pair of electrons $V(x)$ normalized to $V_B$ is shown in Fig. 2 by the blue solid line. We shall use this approximation for studying the phase transition in the classical approach. In a quantum model, we shall use a slightly simpler interaction potential. At $x > a$, it coincides with the form of Eq. (4), takes into account the hard core repulsion at $x < d$, but neglects $e^2/\varepsilon x$ in the region $a > x > d$. The lines of the Fig. 2 were plotted for $\varepsilon = 10$ and $a = 1$ nm. The Fermi energy normalized to $V_B$ is indicated by horizontal lines calculated for the model of spinless fermions using two values of the effective mass: $m^* = m_e$ and $m^* = 0.25 m_e$ (here, $m_e$ is the free electron mass). It should be noted that the lines of the figure remain the same if $\varepsilon$ is changed proportionally to $a$ (for example: $\varepsilon = 5$ and $a = 0.5$ nm).

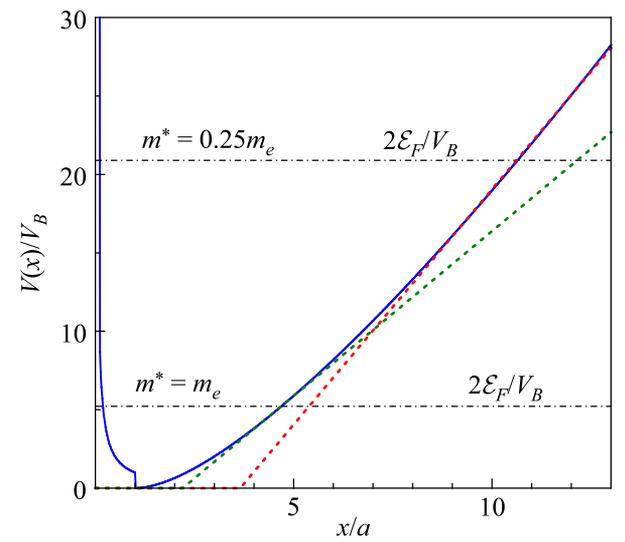

*Fig. 2.* (Color online) The potential energy of the two nearest electrons normalized versus the dimensionless parameter $x/a$ (blue solid line). The horizontal lines (dash-dotted) indicate the maximum kinetic energy of the pair $2\mathcal{E}_F/V_B$ for two different effective masses. Two straight lines (dotted) demonstrate two linear approximations suitable for the chosen effective masses.

Impenetrable fermions are forced to have different momentums, therefore their motion exposes the background. From Fig. 2, it follows that a pair of electrons with opposite momentums and energy $\mathcal{E}_k \simeq \mathcal{E}_F$ can be separated to a large extent because of zero-point vibrations. One can define a typical amplitude of zero-point vibrations (a correlation length) $\xi$ by the condition

$$V(\xi) = 2\mathcal{E}_F.$$

For $\xi \gg a$, sometimes it is useful to neglect the logarithmic dependence in Eq. (5) simplifying:

$$V(x) \simeq V_B \lambda (x/a - \beta) \theta(x/a - \beta), \quad (6)$$

where the numbers $\lambda$ and $\beta$ to be adjusted for better fitting the real line within a chosen range of $\xi$. For example, at $\xi \approx 10a$ a reasonable fitting is obtained by fixing $\lambda = 3$, and $\beta = 3.65$. For $\xi \approx 5a$, there is a better choice: $\lambda = 2.1$ and $\beta = 2.2$, as indicated in Fig. 2 by dotted straight lines. Using the simplification of Eq. (6), $\xi$ can be represented as

$$\xi = \left(\frac{2\mathcal{E}_F}{\lambda V_B} + \beta\right) a. \quad (7)$$

Thus, the amplitude of zero-point vibrations $\xi$ can be much larger than the average distance between electrons $a$. The amplitude $\xi$ can be considered as a typical pairing size of two nearest electrons. Remarkably, the condition $\xi \gg a$ resembles the result of the BCS theory: there are a great many electrons within the Cooper pair radius. However, in our case, there are no other electrons between the nearest electrons. It should be noted that even the formal inclusion of the Fermi pressure in the attractive potential cannot make the correlation length less than three average distances between electrons.

Already at this stage, we can estimate roughly the quasiparticle energy gap using Eq. (7) and the conventional relation between the gap $\Delta(0)$ and the intrinsic coherence length

$$\Delta(0) = \frac{\hbar v_F}{\pi \xi}. \qquad (8)$$

Substituting here $\xi$ from Eq. (7) one can find

$$\Delta(0) = \frac{\lambda V_B}{\pi^2 \left(1 + \lambda \beta \dfrac{V_B}{2\mathcal{E}_F}\right)}. \qquad (9)$$

It is interesting that in the system considered, the quasiparticle gap at zero temperature has a rather weak dependence on the quantum parameter $V_B/2\mathcal{E}_F$ assumed to be small, and $\Delta(0)$ is mostly determined by the Coulombic parameter $V_B = e^2 n_p / \varepsilon$ of the positive background.

For the accurate form of $V(x)$ given in Eqs. (4) and (5), the dependence of $\Delta(0)$ on the quantum parameter $2\mathcal{E}_F/V_B$ is found numerically and shown in Fig. 3 by solid lines for two sets of parameters: $\varepsilon = 10$, $a = 1$ nm (blue solid line) and $\varepsilon = 5$, $a = 0.5$ nm (red solid line), which keep $2\mathcal{E}_F/V_B \propto \varepsilon/a$ the same (here we assume $n_e = n_p$). The approximate form of Eq. (9) is shown by dotted and dashed lines for two sets of $\lambda$ and $\beta$ given above. Vertical lines in Fig. 3 indicate the values of the parameter $2\mathcal{E}_F/V_B$ corresponding to $m^* = m_e$ and $m^* = 0.25 m_e$. The gap values shown in this figure make a detailed study of this electron system very promising.

Thus, a strong electron-electron repulsion at small distances and a rather weak electron attraction to the background charge exposed are reduced to the interaction of an infinite range between impenetrable electrons:

$$U(x_1, x_2, ..., x_N) = \sum_{i=1}^{N-1} V(|x_{i+1} - x_i|), \qquad (10)$$

where $N$ is the total number of electrons. We shall use this form for studying phase transitions in the system. According to the results of the qualitative analysis shown in Fig. 3, at a certain range of the system parameters, these transitions can occur even at room temperature.

### 3. Classical phase transition

It is instructive to consider the properties of a 1D classical gas of impenetrable particles with mutual interaction given by Eqs. (4), (5), and (10), even though the interaction potential is found assuming that $\mathcal{E}_F \gg V_B$. In this Section, the electron density $n_e = 1/a$ is not fixed to the background charge density. Then, assuming the pressure is zero, the average distance between electrons:

$$\langle x_{i+1} - x_i \rangle = \frac{\int_0^\infty x e^{-V(x)/T} dx}{\int_0^\infty e^{-V(x)/T} dx}. \qquad (11)$$

The integrals entering Eq. (11) are finite because $V(x)$ increases with $x$ at $x > a$.

Since the particle density $1/a$ can be varied, while the background density assumed to be fixed $n_p \equiv 1/a_+ = \mathrm{const}$, one should separate $a$ and $a_+$, and rewrite the interaction potential at $x > d$ as

$$V(x) = V_B \left[ \varphi(x/a)\theta(x/a - 1) + \frac{a_+}{x}\theta(a_+ - x) \right]. \qquad (12)$$

The second term in square brackets takes into account the repulsion potential $e^2/\varepsilon x$ at $x < a_+$ discussed above. It is clear that an increase of electron density above the condition $a = a_+$ would increase strongly the system energy.

Then, assuming that $V_d \gg T$ and $a \geq a_+$, the denominator of Eq. (11) can be represented as a linear function of the variable $a$:

$$a_+ \alpha(T) + (a - a_+) + a B(T), \qquad (13)$$

where

$$\alpha(T) = \int_{d/a_+}^1 e^{-\frac{V_B}{T}\frac{1}{y}} dy, \quad B(T) = \int_1^\infty e^{-\frac{V_B}{T}\varphi(y)} dy. \qquad (14)$$

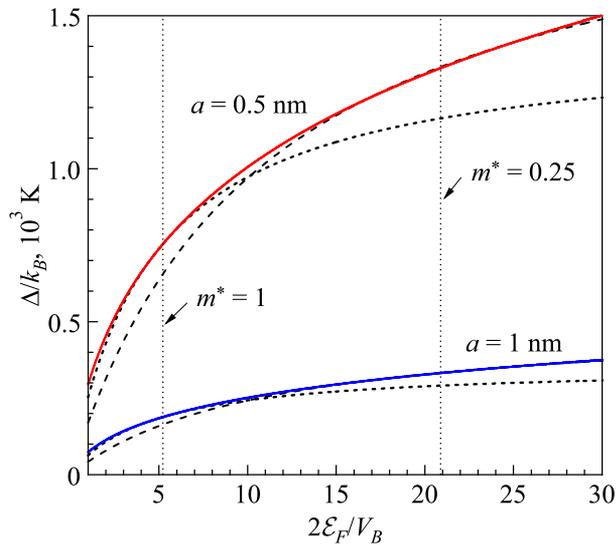

*Fig. 3.* (Color online) The dependence of $\Delta$ on the ratio $2\mathcal{E}_F/V_B$ found from the qualitative analysis based on Eq. (8) and the condition $V(\xi) = 2\mathcal{E}_F$. Solid lines show the result obtained for the accurate form of $V(x)$: $\varepsilon = 10$, $a = 1$ nm (blue solid line) and $\varepsilon = 5$, $a = 0.5$ nm (red solid line). The approximate solution of Eq. (9) is shown by dotted and dashed lines calculated using two sets of parameters $\lambda$ and $\beta$ indicated in the text. The effective mass $m^*$ is indicated in units of $m_e$.

Similarly, the numerator of Eq. (11) can be represented as a quadratic form of the variable $a$:

$$a_+^2 \gamma(T) + \frac{a^2 - a_+^2}{2} + a^2 \left[ B(T) + D(T) \right], \quad (15)$$

where

$$\gamma(T) = \int_{d/a_+}^1 y e^{-\frac{V_B}{T} \frac{1}{y}} dy, \quad D(T) = \int_0^\infty y e^{-\frac{V_B}{T} \varphi(y+1)} dy. \quad (16)$$

The introduced above quantities $\alpha$, $\gamma$, $B$, and $D$ are dimensionless functions of temperature independent of $a$.

In the self-consistent approximation $\langle x_{i+1} - x_i \rangle = a$, we have a quadratic equation for $a(T)$ whose solution can be easily found

$$\frac{a}{a_+} = \frac{(1-\alpha) + \sqrt{(1-\alpha)^2 - (1-2D)(1-2\gamma)}}{(1-2D)}. \quad (17)$$

Another solution of the quadratic equation results in $a(T)$ decreasing with $T$, and, therefore, should be rejected. The quantity $D(T)$ increases steadily with temperature. From Eq. (17), one can see that there is a critical temperature defined by the condition $2D(T_c) = 1$. According to Eq. (16), $T_c$ depends only on the properties of the potential $V(x)$ at large distances. The numerical solution gives $T_c \simeq 0.59 V_B$.

If one takes into account that at $T \leq T_c$ the numerical parameters $\alpha \leq 0.056$ and $\gamma \leq 0.044$ are very small, the solution of Eq. (17) can be substantially simplified:

$$\frac{a}{a_+} \simeq \frac{1}{1-\sqrt{2D}}. \quad (18)$$

The exact form and the approximation of Eq. (18) are shown in Fig. 4. At low temperatures $T \ll T_c$ the electron spacing $a$ is very close to $a_+$, but it increases fast at $T \lesssim T_c$.

It is interesting to note that at the zero-pressure condition, the potential given in Eq. (12) allows uniform expanding $a \to \infty$ without an increase in $V(x)$. This is the reason why the expansion of $a$ appears at $T \to T_c$ despite the infinite-range attraction. Thus, at $T > T_c$, we have a gas phase of the system because one must apply pressure to realize a finite density. In the opposite range $T < T_c$, the electron system has a finite spacing, which means that it is in a condensed phase. The critical temperature for this phase transition depends only on the Coulomb parameter of the neutralizing background $T_c \simeq 0.59 V_B = 0.59 e^2 / \varepsilon a_+$. It is substantially higher (approximately 5 times) than the critical temperature for the superconducting transition expected from Fig. 3 assuming $T_c = \Delta / 2$. Anyway, in the following, we shall assume that the average electron density coincides with the background density and $a = a_+$.

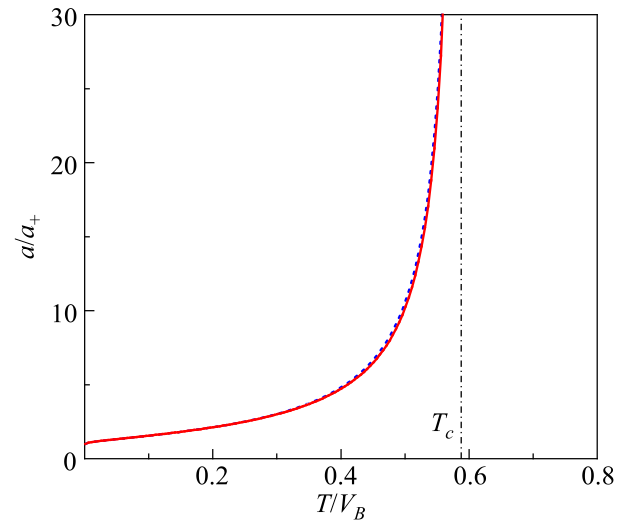

*Fig. 4.* (Color online) The electron spacing $a$ normalized versus temperature in an infinite 1D classical system of impenetrable particles interacting via the potential of Eq. (12). The red solid line represents the accurate solution given in Eq. (17), while the blue dotted line indicates the approximate solution given in Eq. (18).

### 4. Quantum regime. Modeling of the superconducting state

Considering a quasi-1D system of strongly interacting particles ($V_d = e^2 / d \gg \mathcal{E}_F$) with the effective attractive potential between nearest neighbors of infinite range [$V(x) \propto x$] one cannot proceed without substantial simplifications. Assume first that the impenetrable core is much smaller than the particle spacing $d \ll a$. In this case, it is reasonable to analyze the limiting case when the impenetrable core has shrunk to a point and $V_d \to \infty$. Then, the condition induced by the core is reduced to

$$\psi(x_1, ..., x_N) = 0, \quad \text{if} \quad x_j = x_{j+1}. \quad (19)$$

It is well-known [5] that for a system of spinless fermions, point interaction can be neglected because the Fermi wave functions automatically vanish when two particles come together. Even impenetrable bosons can be described in terms of the free spinless Fermi gas eigenfunctions [5].

That is, the impenetrable core acts as a stability factor. Therefore, when constructing a quantum model of 1D impenetrable electrons, it is reasonable to start with the general Hamiltonian of spinless fermions

$$H = \sum_k \varepsilon_k c_k^\dagger c_k + \frac{1}{2L} \sum_{q,k,k'} \frac{V_q}{N} c_{k+q}^\dagger c_{k'-q}^\dagger c_{k'} c_k, \quad (20)$$

where $\varepsilon_k = \mathcal{E}_k - \mathcal{E}_F$ and $\mathcal{E}_k = \hbar^2 k^2 / 2m^*$, $c_k^\dagger$ and $c_k$ are creation and destruction operators that obey anticommutation relations, $L$ is the size of the 1D system, and $V_q$ is the Fourier-transform of the pair-interaction potential $V(x)$. In the denominator of the interaction term, the number of electrons $N$ appears because the total potential energy given in Eq. (10) contains only one sum over all particles, instead of two sums. Qualitatively, its appearance is similar to the appearance of the number 2 in the same term which takes into account double counting. Anyway, we intend to construct a model of spinless fermions with infinitesimal pair interaction of infinite range, and Eq. (20) satisfies our goal. It should be noted that the factor $1/N$ makes the interaction potential infinitesimal only in the many-electron sense.

Another essential point of our model Hamiltonian is the Fourier transform of the interaction potential. For the potential given above in Eqs. (4) and (5) we have to consider the properties of the Fourier transform:

$$V_q = 2V_B a \int_1^\infty \varphi(\rho)\cos(qa\rho)\,d\rho, \qquad (21)$$

where $\varphi(\rho)$ is from Eq. (5). It is instructive to analyze first the approximate form of Eq. (6)

$$\varphi(\rho) \approx \lambda(\rho-\beta)\theta(\rho-\beta).$$

In this case, similar to the Fourier-transform of the usual 3D Coulomb potential, the integral of Eq. (21) is not well defined, since it oscillates fast at infinity. Following the usual procedure [7], we assume these oscillations damp out. Therefore, in the limiting case $qa \ll 1$ important for the following treatment, we have

$$V_q \simeq -\frac{2\lambda V_B}{a}\frac{1}{q^2}. \qquad (22)$$

Thus, Eq. (22) is similar to the Fourier transform of the usual 3D Coulomb potential $|V_q| \propto q^{-2}$, but it has the negative sign. The approximation of Eq. (22) with a fixed $\lambda$ was used in our short note [4]. For the accurate form of $\varphi(\rho)$, we also shall use only the low-wavelength asymptote

$$V_q \simeq -\frac{2V_B}{a}\frac{1}{q^2}\ln\!\left(\frac{2}{|q|a}\right), \qquad (23)$$

which has an additional logarithmic factor important at $qa \ll 1$.

The background charge becomes exposed to the largest extent when electrons of a pair have opposite momenta. Therefore, the next approximation is to reduce the interaction term of the Hamiltonian

$$H = \sum_{k>0}\varepsilon_k\left(c^\dagger_k c_k + c^\dagger_{-k} c_{-k}\right)$$
$$+ \frac{1}{L}\sum_{k>0,k'>0}\frac{V_{k'-k}}{N} c^\dagger_k c^\dagger_{-k'} c_{-k'} c_k. \qquad (24)$$

Here, the double sum over momentums $k$ and $k'$ is transformed to positive values of summation indexes, and we neglected terms containing $V_{k'+k}$ because only low values of $|q|$ are important due to Eqs. (22) and (23).

The Hamiltonian of Eq. (24) has the form that allows applying the standard methods of the theory of superconductivity [8]. Since there are many ways to obtain the same results, we shall choose the simplest one. Firstly, we assume that in the new ground state, the *c*-numbers $b_k = \langle c_{-k} c_k \rangle$ and $b^*_k = \langle c^\dagger_k c^\dagger_{-k} \rangle$ are finite. Then, the differences $c_{-k} c_k - b_k$ and $c^\dagger_{k'} c^\dagger_{-k'} - b^*_{k'}$ are assumed small. Finally, the Hamiltonian of Eq. (24) can be reduced to the canonical form using the unitary transformation:

$$c_k = u_k A_{1,k} + v_k A^\dagger_{2,k}, \qquad c_{-k} = u_k A_{2,k} - v_k A^\dagger_{1,k}, \qquad (25)$$

where $A_{1,k}$, $A^\dagger_{1,k}$, and $A_{2,k}$, $A^\dagger_{2,k}$ are two new species of fermion operators.

The transformation parameters $u_k$ and $v_k$ (normally referred to as coherence factors) are defined using the conventional procedure, which gives

$$u^2_k = \frac{1}{2}\!\left(1+\frac{\varepsilon_k}{E_k}\right), \qquad v^2_k = \frac{1}{2}\!\left(1-\frac{\varepsilon_k}{E_k}\right), \qquad (26)$$

where

$$E_k = \sqrt{\varepsilon^2_k + \Delta^2_k}, \qquad \Delta_k = -\frac{1}{L}\sum_{k'>0}\frac{V_{|k'-k|}}{N} b_{k'} \qquad (27)$$

are the quasiparticle energy and the superconducting gap.

Similar to the conventional theory, the new ground state is defined as

$$|0\rangle_{\text{new}} = \prod_{k>0}\!\left(u_k + v_k c^\dagger_k c^\dagger_{-k}\right)|0\rangle. \qquad (28)$$

It obeys the conditions $A_{1,q}|0\rangle_{\text{new}} = 0$ and $A_{2,q}|0\rangle_{\text{new}} = 0$, where

$$A_{1,k} = u_k c_k - v_k c^\dagger_{-k}, \qquad A_{2,k} = v_k c^\dagger_k + u_k c_{-k}. \qquad (29)$$

Using Eq. (29), one can also find

$$b_k = \frac{\Delta_k}{2E_k}\tanh\!\left(\frac{E_k}{2T}\right), \qquad (30)$$

and

$$\Delta_k = -\frac{1}{LN}\sum_{k'>0}V_{k'-k}\frac{\Delta_{k'}}{2E_{k'}}\tanh\!\left(\frac{E_{k'}}{2T}\right). \qquad (31)$$

The last equation determines the superconducting gap $\Delta_k$. An important difference of the new gap equation, as compared to the conventional BCS theory, is that here the attractive potential $V_q$ is not restricted to a narrow range near the Fermi energy. Another important difference is the presence of the $1/N$ factor which will compensate for the divergence caused by the attraction of infinite range.

In contrast to the conventional theory, the Fourier-transform of the interaction potential $V_q \propto -q^{-2}$ is singular at small $q$, which, together with the above-noted important differences, leads to a large gap at the Fermi level without any small exponential factor. In this case, the main contribution to the sum of Eq. (31) comes from $|k'-k| \equiv |q| \ll k_F$, and, therefore, we can replace $\Delta_{k+q}$ and $E_{k+q}$ with $\Delta_k$ and $E_k$, respectively. This approximation gives

$$1 = \frac{\Delta_F}{\sqrt{\varepsilon_k^2 + \Delta_k^2}} \tanh\left(\frac{\sqrt{\varepsilon_k^2 + \Delta_k^2}}{2T}\right), \quad (32)$$

where

$$\Delta_F = \frac{V_B}{2\pi^2}\left(\ln\frac{L}{\pi a} - 1\right), \quad (33)$$

and the smallest $|q|$ is set to $2\pi/L$. For the approximate form of $\varphi(\rho)$ given in Eq. (6), one can find

$$\Delta_F \simeq \frac{\lambda V_B}{2\pi^2}, \quad (34)$$

which is quite close to the result of the qualitative analysis given above [Eq. (9)] and considered in the limiting case $V_B/2\mathcal{E}_F \ll 1$. For the formal inclusion of the Fermi pressure discussed above, in Eq. (34), one should use the replacement $\lambda V_B \to 2\mathcal{E}_F/3$ which gives $\xi = 6a$, where the correlation length $\xi$ is defined by Eq. (8).

According to Eq. (32), $\Delta_F$ represents the superconducting gap at the Fermi level $k = k_F$ when $T \to 0$. The gap $\Delta_k$ should depend on $k$ in such a way that $E_k = \sqrt{\varepsilon_k^2 + \Delta_k^2}$ remains constant until $\Delta_k^2$ becomes zero. Specifically,

$$\Delta_k = \sqrt{\Delta_F^2 - \varepsilon_k^2}, \quad (35)$$

and it differs from zero only in the vicinity of the Fermi level $k_- < k < k_+$, here

$$k_\pm = k_F\sqrt{1 \pm \frac{\Delta_F}{\mathcal{E}_F}}. \quad (36)$$

The respective region exists also for negative $k$. It is remarkable that in contrast with the conventional BCS theory, in the model considered here, the gap exists only in a narrow region near the Fermi energy even though the attraction potential is not restricted to this region. The comparison between typical quasiparticle spectra of the BCS theory and 1D impenetrable electrons (IE) is given in Fig. 5. Here, the BCS gap is set to $\Delta_F$, though its real value is usually much smaller than that found here.

Consider now the temperature dependence of the superconducting gap at the Fermi level $\Delta_{k_F}(T) \equiv \Delta_F \varkappa(T)$.

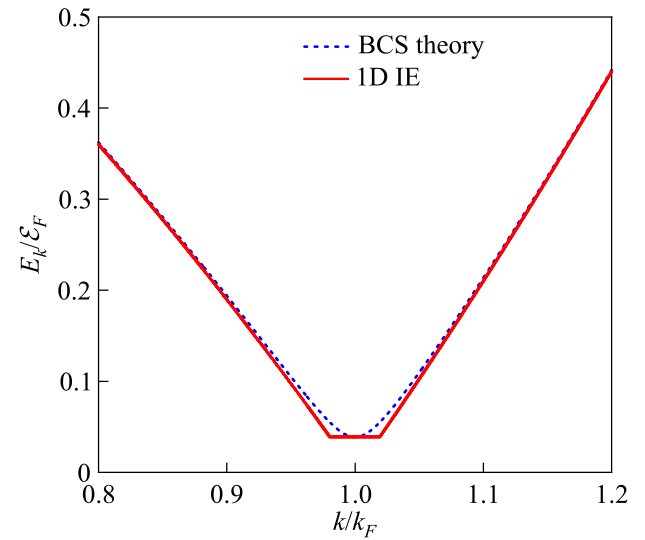

*Fig. 5.* (Color online) The quasiparticle energy normalized versus $k/k_F$ for the model of spinless fermions with infinitesimal attraction of infinite range applied to the 1D impenetrable-electron (IE) system (red solid line). The corresponding result of the BCS theory with the same superconducting gap at the Fermi level $\Delta_k = \Delta_F$ is shown by the blue dotted line.

Here, the dimensionless function $\varkappa$ is determined by the transcendental equation:

$$\varkappa = \tanh\left(\frac{\Delta_F}{2T}\varkappa\right), \quad (37)$$

which is well-known in statistical physics. A nonzero solution of Eq. (37) exists only at $\Delta_F/2T > 1$. Therefore, $T_c = \Delta_F/2$, which is not far from the BCS relationship $T_c \simeq 0.568\Delta(0)$. The temperature dependence of the dimensionless parameter $\varkappa$ can be easily found from Eq. (37) by numerical evaluation. Here, we give only a simple analytical approximation

$$\varkappa \equiv \frac{\Delta_{k_F}}{\Delta_F} \simeq \sqrt{1 - \left(\frac{T}{T_c}\right)^{3.2}}, \quad (38)$$

which is quite close to the numerical solution near the critical temperature.

### 5. Discussion and conclusion

Thus, the model of spinless fermions with infinitesimal attraction of infinite range allows us to take into account strong repulsion acting between electrons at small distances ($V_d \to \infty$) and avoid problems with the singular nature of the Fourier transform of effective attraction induced by electron-electron correlations. It is important that the superconducting gap obtained in the framework of this model is remarkably similar to that found in Sec. 2 using a simple qualitative analysis based on the linear approximation $\varphi(\rho) \approx \lambda(\rho - \beta)$ and Eq. (8) for the intrinsic coherence length $\xi$.

For example, compare Eq. (34) with Eq. (9) obtained in the qualitative analysis. In the limiting case $2\mathcal{E}_F/V_B \to \infty$, the strict solution of the model given in Eq. (34) differs from Eq. (9) only by the additional number 2 in the denominator. Remarkably, at finite values of the parameter $2\mathcal{E}_F/V_B$ indicated in Fig. 3, within the range 5–30 the expression in brackets of the denominator of Eq. (9) varies between 1.92 and 1.55 which is not far from 2. Therefore, keeping in mind the relationship $T_c = \Delta_F/2$, qualitative results shown in Fig. 3 can be used to estimate the critical temperature for given values of the system parameters ($m^*$, $a$, and $\varepsilon$).

Here, we considered only the limiting case $V_d \to \infty$ which makes electrons impenetrable. The case of finite $V_d$ requires special considerations. Anyway, the estimates given here for typical system parameters indicate that the ratio $V_d/\mathcal{E}_F$ could be about $10$–$10^2$ which justifies the approximation of impenetrable particles.

It is well-known that an interacting 1D fermion gas of penetrable electrons is unstable against the formation of a charge density wave (CDW) with wavevector $2k_F$ (the so-called Peierls instability). In the mean-field theory, the corresponding order parameter $\Delta_{\text{CDW}}$ is defined as the expectation value of another pair of fermion operators. We expect that the stability factor $V_d/\mathcal{E}_F \gg 1$ discussed above would favor the spinless fermion state and the superconducting transition. The main reason for that is the fact that we chose the expectation value of the fermion operators, $\langle c_{-k} c_k \rangle$, which leads to the maximum contribution from $V_q < 0$ into the order parameter [see Eqs. (27) and (31)]. Indeed, besides the strong reduction of the expectation value of Coulomb repulsion at $x \sim d$, when comparing the total energy gain due to the superconducting transition and the CDW transition, we suppose that abnormally high values of the superconducting gap shown in Fig. 3 will also make the superconducting transition more favorable.

Another approximation used here concerns disregarding the effective attraction between electrons which are not nearest neighbors. For example, when two electrons with opposite momenta have few electrons in between them ($N_*$), at a large distance the electrons swept are attracted to the partly exposed positive background whose charge is reduced by the number of remaining electrons $N_*$. We expect that this effect would only increase the quasiparticle gap found here.

Interestingly, several different superconductivity mechanisms somehow relate to the screening effect. In the BCS theory, electron screening of the ion lattice disturbance reduces the effective attraction responsible for Cooper pairing. In another mechanism present even in the case of purely repulsive forces between the particles [3], the quantum effects of electron screening are responsible for superconductivity. A sharp Fermi surface is necessary for the effect in this theory. In the theory discussed here, the superconducting transition occurs due to the lack of screening of the rigid neutralizing background exposed by a motion of impenetrable electrons in a quasi-1D system. For this case, according to Eqs. (31) and (32) a sharpness of Fermi distribution seemingly is not necessary.

Concluding, in this work we described the properties of a quasi-1D electron system on a rigid, uniform, neutralizing background under conditions where it can be considered as a quantum liquid of impenetrable charged particles. The advantage of such a liquid state is that it eliminates screening of the background charge exposed by electron motion, which results in an effective attraction between nearest electrons of infinite range. This kind of interaction can lead to a phase transition even in the 1D system. Moreover, in the state described here, all electrons are involved in the long-range pairing, which is in contrast with the conventional theory. To describe the superconducting transition we used the model of spinless fermions with infinitesimal attraction of infinite range, which leads to results that are very close to that obtained here using the simple qualitative analysis. Due to the singular nature of the attractive potential, the superconducting gap and the critical temperature found have no exponentially small factor, and, for typical values of the system parameters, they can be above room temperature.

The author thanks V. V. Slavin, S. I. Shevchenko, and A. A. Zvyagin for helpful discussions.

1. A. Mourachkine, *Room-Temperature Superconductivity*, Cambridge International Science Publishing, United Kingdom (2004) [ISBN: 1-904602-27-4].
2. F. Altomare and A. M. Chang, *One-Dimensional Superconductivity in Nanowires*, WILEY-VCH Verlag GmbH & Co. KGaA, Weinheim, Germany (2013) [ISBN 978-3-527-64907-5].
3. W. Kohn and J. M. Luttinger, *Phys. Rev. Lett*. **15**, 524 (1965).
4. Yu. P. Monarkha, *Fiz. Nizk. Temp*. **15**, 1204 (1989) [*Sov. Low Temp. Phys.* **15**, 664 (1989)].
5. M. Girardeau, *J. Math. Phys.* **1**, 516 (1960).
6. T. Giamarchi, *Quantum Physics in One Dimension*, Clarendon Press & Oxford, New York (2003).
7. G. D. Mahan, *Many-Particle Physics*, Kluwer Academic / Plenum Publishers, 3rd ed., New York, London (2000), p. 19.
8. K. Fossheim and A. Sudbo, *Superconductivity Physics and Applications*, John Wiley & Sons Ltd, England (2004) [ISBN 0-470-84452-3].

Електрон-електронне притягання, яке обумовлене кулонівськими кореляціями, та можлива надпровідність в одновимірній електронній рідині на жорсткому нейтралізуючому фоні

Yu. P. Monarkha

Теоретично проаналізовано умови, за яких квазіодновимірну (1D) електронну систему можна розглядати як квантову рідину з непроникних заряджених частинок. Показано, що при наявності інертного, нейтралізуючого фону рух непроникних електронів оголяє позитивний заряд, що призводить до ефективного взаємного притягання з нескінченно великим радіусом дії. Як наслідок, усі електрони беруть участь у далекодіючому спарюванні. Для опису спектра збуджень та надпровідної щілини в 1D електронних каналах низької густини запропоновано модель безспінових ферміонів з нескінченно слабким притяганням нескінченного радіуса дії. На відміну від традиційної теорії енергетична щілина не містить експоненційно малих факторів. Вона переважно залежить від кулонівських параметрів системи, що дає орієнтири щодо практичних аспектів високотемпературної надпровідності.

Ключові слова: одновимірний електронний газ, електронні кореляції, надпровідна щілина.